\documentclass[aps,prl,reprint,groupedaddress,amsmath,amssymb,superscriptaddress]{revtex4-1}
\usepackage{graphicx}

\usepackage{footmisc}

\usepackage[usenames,dvipsnames]{color}

\begin{document}
\title{Optimal Electromagnetic Searches for Axion and Hidden-Photon Dark Matter}

\author{Saptarshi Chaudhuri}
\affiliation{Department of Physics, Stanford University, Stanford, CA 94305}

\author{Kent D. Irwin}
\affiliation{Department of Physics, Stanford University, Stanford, CA 94305}
\affiliation{Kavli Institute for Particle Astrophysics and Cosmology, Stanford University, Stanford, CA 94305}
\affiliation{SLAC National Accelerator Laboratory, Menlo Park, CA 94025}

\author{Peter W. Graham}
\affiliation{Department of Physics, Stanford University, Stanford, CA 94305}
\affiliation{Kavli Institute for Particle Astrophysics and Cosmology, Stanford University, Stanford, CA 94305}
\affiliation{Stanford Institute for Theoretical Physics, Department of Physics, Stanford University, Stanford, CA 94305}

\author{Jeremy Mardon}
\affiliation{Department of Physics, Stanford University, Stanford, CA 94305}
\affiliation{Stanford Institute for Theoretical Physics, Department of Physics, Stanford University, Stanford, CA 94305}

\date{\today}

\begin{abstract}

Direct-detection searches for axions and hidden photons are playing an increasingly prominent role in the search for dark matter. In this work, we derive the properties of optimal electromagnetic searches for these candidates, subject to the Standard Quantum Limit (SQL) on amplification. We show that a single-pole resonant search may possess substantial sensitivity outside of the resonator bandwidth and that optimizing this sensitivity may increase scan rates by up to five orders of magnitude at low frequencies. Additional enhancements can be obtained with resonator quality factors exceeding one million, which corresponds to the linewidth of the dark matter signal. We present the resonator optimization in the broader context of determining the optimal receiver architecture (resonant or otherwise). We discuss prior probabilities on the dark matter signal and their role in the search optimization. We determine frequency-integrated sensitivity to be the figure of merit in a wideband search and demonstrate that it is limited by the Bode-Fano criterion. The optimized single-pole resonator is approximately 75\% of the Bode-Fano limit, establishing it as a fundamentally near-ideal, single-moded dark matter detection scheme. Our analysis shows, in contrast to previous work, that the scanned single-pole resonant search is superior to a reactive broadband search. Our results motivate the broad application of quantum measurement techniques evading the SQL in future axion- and hidden-photon dark matter searches.
\end{abstract}

\maketitle

The existence of dark matter is one of the most compelling pieces of evidence for physics beyond the Standard Model.\cite{ade2014planck} Two of the leading candidates for dark-matter particles are weakly interacting massive particles (WIMPs) and axions.\cite{abbott1983cosmological,Dine:1982ah,Preskill:1982cy,takahashi2018qcd,graham2018stochastic} The axion is a pseudoscalar originally motivated as a solution to the strong CP problem \cite{peccei1977cp,weinberg1978new,wilczek1978problem}. Because of their low (usually sub-eV) mass, and therefore high number density if they make up a majority of the dark matter, axions are best described as classical fields oscillating at a frequency slightly higher than their rest frequency $\nu_{\rm DM}^{0}=m_{\rm DM}c^{2}/h$, where $m_{\rm DM}$ is the rest mass. The observed frequency is slightly higher than the rest-mass frequency because of the $\sim 10^{-3}$c virial velocity in the Milky Way; the resulting spread in kinetic energy gives axion dark matter a bandwidth of $\Delta \nu_{\rm DM} \sim 10^{-6} \nu_{\rm DM}^{0}$. The objective of axion searches is to detect this narrowband signal through its coupling to the Standard Model, either through the strong force \cite{Budker:2013hfa} or electromagnetism \cite{Sik83,Sik85}. We focus on electromagnetic detection, in which an axion converts to a photon in a background electromagnetic field. 

The hidden photon, a vector, is another well-motivated field-like dark-matter candidate that couples to electromagnetism (via kinetic mixing with the visible photon). It emerges generically in Standard Model extensions\cite{holdom1986searching,nelson:2011sf,arias2012wispy} and may be produced by cosmic inflation.\cite{Graham:2015rva}

To date, most direct-detection searches for dark matter have focused on WIMPs.\cite{Agn15,Ake17,aprile2012dark} For axions and hidden photons, a wide range of masses and couplings (the axion-photon coupling $g_{a\gamma\gamma}$ or the hidden photon mixing angle $\varepsilon$) have not been explored. Accordingly, numerous searches looking for the conversion to photons have recently been proposed.\cite{graham2015experimental,irastorza2018new}

Because of the interest in these candidates, it is important to conduct a broad first-principles optimization of search sensitivity for probes of the electromagnetic coupling. A number of searches are based on tunable resonant structures.\cite{du2018search,braine2020extended,brubaker2017first,zhong2018results,brubaker2017haystac,Chaudhuri:2014dla,Silva-Feaver:2016qhh,phipps2020exclusion,Sik14,caldwell2017dielectric} When the resonant frequency is tuned to $\sim\nu_{\rm DM}^{0}$, the signal induced by dark matter is enhanced in the detector. One can conduct a sensitive search for dark matter by tuning the resonator across the desired search range. Are single-pole resonators optimal?  Or can better sensitivity, integrated across a search band, be obtained with a more broadband structure, such as an LR circuit \cite{kahn2016broadband,ouellet2019first,salemi2021search} or a free-space antenna \cite{horns2013searching}? 

The principal purpose of this paper is to determine the optimal characteristics of a dark matter receiver, subject to the Standard Quantum Limit (SQL) on phase-insensitive amplification.\cite{caves1982quantum} The primary conclusion is that the optimized single-pole resonator is a fundamentally near-ideal, single-moded, linear, passive detector for probing dark matter across a wide band. We derive the optimized resonant scan, which exploits the substantial sensitivity available outside the resonator bandwidth. This optimization can increase scan rates by up to five orders of magnitude at low frequency. We further show that scan rate is enhanced by higher resonator quality factor, even for quality factors exceeding one million, which corresponds to the dark matter bandwidth. Though we describe our results specifically for background DC magnetic fields, our primary conclusions for search optimization are the same for background AC fields.\cite{melissinos2008search,sikivie2010superconducting,Berlin:2019ahk} See \cite{chaudhuri2018fundamental} for an extended discussion of the results presented here. 

In order to optimize the receiver, we consider the signal-to-noise ratio (SNR) in the detector, taking into account the impedance match to dark matter, the noise match to the amplifier, and periodically-varied receiver parameters (for the scan). We also take into account prior information on the dark matter signal, which can take the form of previous constraints, preferred search ranges, or candidate signals. We optimize a value functional, which is the expectation value of the SNR squared (see eq. 164 of \cite{chaudhuri2018fundamental}) :
\begin{equation}\label{eq:MN_value}
F[H(\nu)]=E[SNR^{2}(H(\nu)); \nu_{\rm DM}^{0}, g_{a\gamma\gamma}/\varepsilon, Q(\vec{v})],
\end{equation}
where $H(\nu)$ is a responsivity function describing how dark matter at frequency $\nu$ excites the detector. The expectation value is taken with respect to user-defined prior probability distributions over the dark matter mass $m_{\rm DM}$ (corresponding to frequency $\nu_{\rm DM}^{0}$), coupling $g_{a\gamma \gamma}$ or $\varepsilon$, and velocity distribution $Q(\vec{v})$. The frequency $\nu$ lies within the dark matter bandwidth, slightly above $\nu_{\rm DM}^{0}$.

\begin{figure}[htp] 
\includegraphics[width=8.6cm]{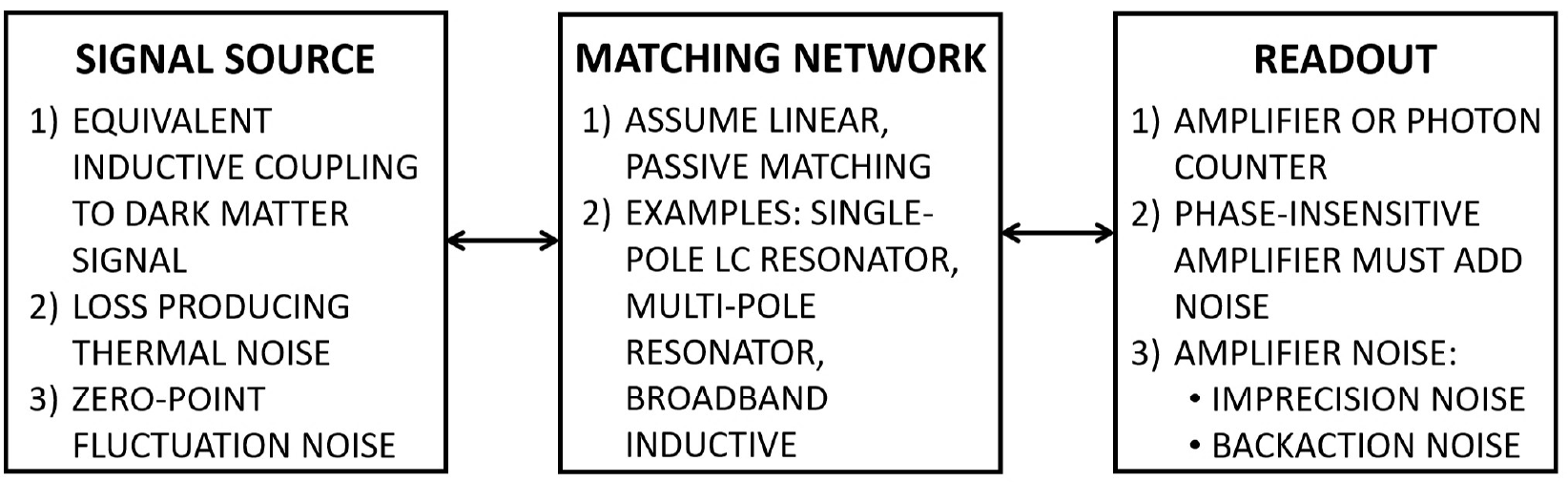}
\caption{Elements of a dark-matter receiver: the signal source, the matching network, and the readout. Double arrows signify that signals travel in both directions through the receiver. \label{fig:DetectorBlockDiagram}}
\end{figure}

We first develop an equivalent circuit to evaluate $H(\nu)$ and receiver SNR and then return to priors. The basic structure of a receiver is displayed in Fig. \ref{fig:DetectorBlockDiagram}. The receiver is optimized by simultaneously optimizing each block and interactions across blocks, subject to priors. We restrict our attention to linear, passive receivers.

The first element of the receiver is the signal source, which contains an element coupling to the electromagnetic fields induced by dark matter. From conservation of energy statements, one can show that there are two categories of coupling: radiative, e.g. through free-space radiation modes (as in a free-space antenna), and reactive, e.g. through energy-storing elements (as in a resonant cavity).  Furtheremore, there are two types of reactive coupling: inductive, through magnetic energy storage, and capactive, through electric energy storage.


In Sec. II and III of \cite{chaudhuri2018fundamental}, we show that radiative coupling is generally disadvantaged relative to reactive coupling, owing to the mismatch between the dark-matter source impedance and the free-space impedance. We thus consider single-moded reactive couplings, which constitute the vast majority of present searches\cite{sikivie2021invisible}. Inductive coupling is always superior to capactive coupling in the regime where the experimental apparatus is smaller than the dark-matter Compton wavelength $\lambda_{\rm DM}^{0}=c/\nu_{\rm DM}^{0}$. When the apparatus is comparable in size to or larger than $\lambda_{\rm DM}^{0}$, inductive and capacitive couplings give comparable sensitivities. 
In the particular case of a cavity resonator, we can model a single mode as an equivalent lumped-LC circuit. Then, coupling to the cavity mode through both the electric and magnetic fields can be modeled as an effective coupling to either the inductor or capacitor. Because of this equivalency, we may, without loss of generality, consider an equivalent circuit with only inductive coupling. 

Intrinsic to every inductive coupling element $L$ is some loss $R$, which produces thermal noise. We thus model the signal source as an equivalent series-$LR$ network. 

The second element of the receiver is an impedance-matching network. It may be used to improve the match to dark matter as well as the match to the readout. For example, the network may be a single-pole resonator (formed by inserting an equivalent-circuit capacitor), an appropriate model for cavity searches ADMX\cite{du2018search} and HAYSTAC\cite{brubaker2017first} as well as lumped-element searches DM Radio\cite{Chaudhuri:2014dla,Silva-Feaver:2016qhh} and ADMX LC\cite{Sik14}. One may also consider using a multi-pole resonator. Yet another architecture is a broadband pickup, for which the $LR$ signal source is connected directly to the input of a SQUID amplifier, e.g. as used in ABRACADABRA and SHAFT.\cite{kahn2016broadband,gramolin2021search}

The third and final element of the receiver is a readout element, which is generically some amplifier or photon counter. In this work, we consider readout with a phase-insensitive amplifier, which amplifies both signal quadratures with equal gain. Such amplifiers are subject to the SQL \cite{caves1982quantum,clerk2004quantum}, which dictates that the measurement must add at least one photon of noise power per unit bandwidth. We utilize the convention that one-half photon of the quantum noise comes from zero-point fluctuations, while the other half comes from the amplifier's two effective noise modes, backaction and imprecision.\cite{clerk2010introduction} 

We now consider how to calculate detector SNR. Probing new parameter space at a particular frequency typically requires integration time much longer than the coherence time $(\Delta \nu_{\rm DM})^{-1}$. In this regime, the signal to be detected is an incoherent power signal, for which the SNR is dictated by the Dicke radiometer equation\cite{dicke1946measurement,asztalos2010squid}:
\begin{equation}\label{eq:Dicke}
SNR=\frac{P_{\rm sig}}{P_{\rm noise}} \sqrt{\Delta \nu_{s} \cdot t}=\frac{P_{\rm sig}}{kT_{S} \Delta \nu_{s}} \sqrt{\Delta \nu_{s} \cdot t},
\end{equation}
where $P_{\rm sig}$ ($P_{\rm noise}=kT_{S}\Delta \nu_{s}$) is signal (noise) power, $T_{S}$ is system noise temperature (a function of the physical and amplifier noise temperatures), $\Delta \nu_{s}$ is effective dark-matter signal bandwidth when convolved with the responsivity function $H(\nu)$ (see Secs. IV and VI of \cite{chaudhuri2018fundamental}), and $t$ is integration time. The system noise temperature is dependent upon the matching network, because the noise temperature of the amplifier depends on the impedance coupled to its input, achieving a minimum when the impedance is ``noise matched.'' In order to fully optimize the third block (and the SNR), we must then simultaneously optimize the second block.

\begin{figure}[htp] 
\includegraphics[width=8.6cm]{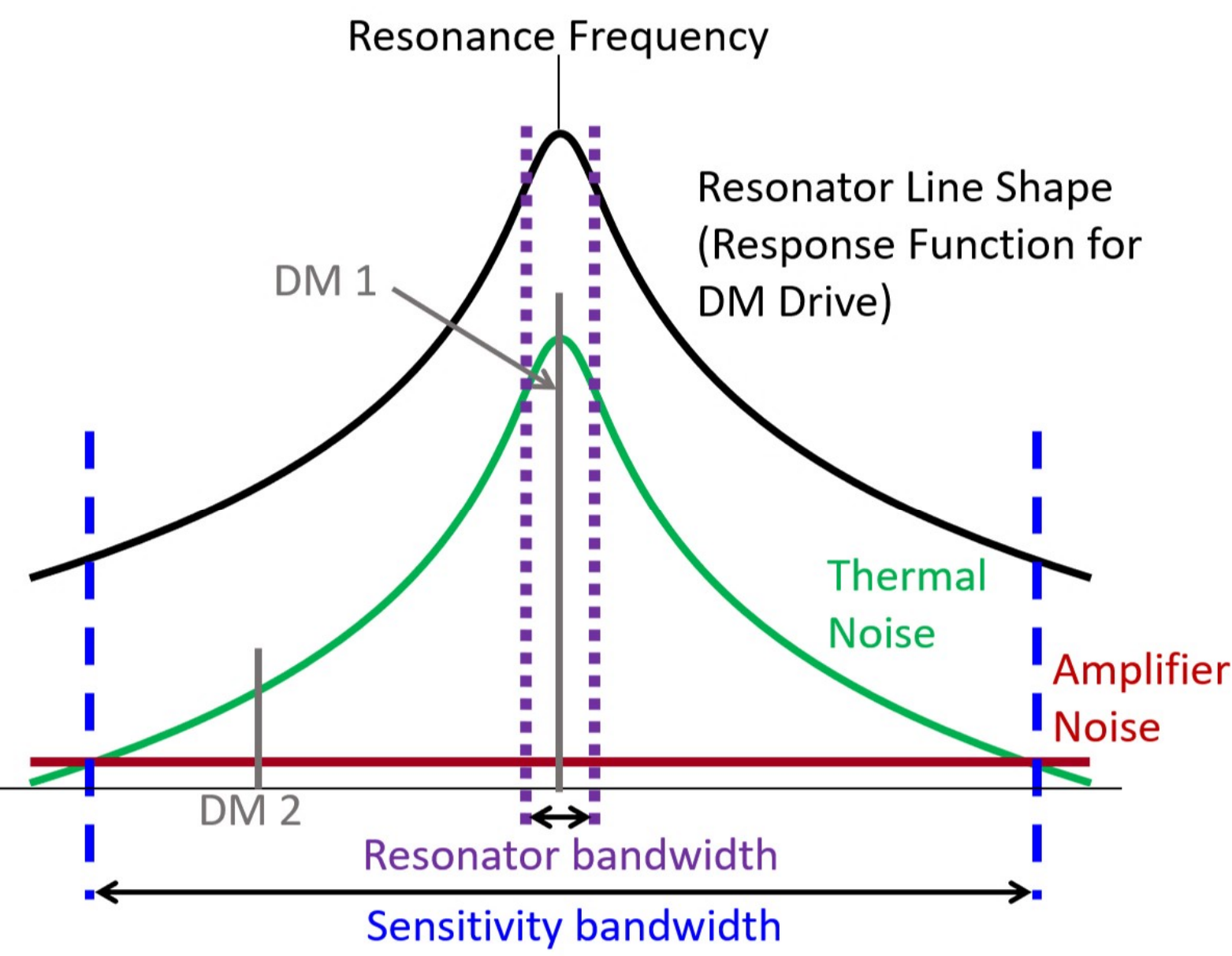}
\caption{Signal and noise in a resonator. The x-axis is frequency detuning (arb. units, linear scale), while the y-axis is power referred to the amplifier input (arb. units, log scale). Green: thermal noise, red: amplifier noise, black: resonator line shape. On- and off-resonance dark matter (DM) signals of equal strength are denoted ``DM 1'' and ``DM 2,'' respectively. Resonator bandwidth is bounded by the dotted purple lines. Sensitivity bandwidth is bounded by dashed blue lines. \label{fig:SensitivityBandwidth}}
\end{figure}

It is in the second block where priors play a central role. Most conceivable priors distribute the mass probability over a wide band. The value functional, eq. (\ref{eq:MN_value}), is then a measure of a detector's frequency-integrated sensitivity. The significance of this observation can be understood by considering a single-pole resonator. See Fig. \ref{fig:SensitivityBandwidth}. The dark-matter signal and thermal noise are both rolled off by the resonator response. As such, the resonator sensitivity to dark matter (i.e. the SNR) remains constant well beyond the resonator bandwidth, as long as amplifier noise is subdominant. For instance, the resonator sensitivity to the on-resonance dark matter signal (``DM 1''), is approximately equal to the sensitivity to a signal of equal strength off-resonance (``DM 2''). The ``sensitivity bandwidth'' is the range over which the amplifier noise is subdominant to thermal noise. This bandwidth can be increased by strengthening the coupling to the amplifier until amplifier backaction (not represented in figure), which degrades SNR, becomes comparable to the thermal noise. With a widely distributed mass prior, we must then choose the matching network not to maximize the on-resonance SNR, but to maximize a weighted product of sensitivity bandwidth and in-band SNR, i.e. integrated sensitivity. The sensitivity bandwidth increases with decreasing amplifier noise. Thus, for optimization, we require quantum-limited readout (achieving the SQL when noise-matched), even when the resonator, of frequency $\nu_{r}$, possesses high thermal occupancy, $h\nu_{r} \ll kT$.

Frequency-integrated sensitivity as the figure of merit for a matching network (resonant or otherwise) is formalized mathematically in terms of priors in Sec. V of \cite{chaudhuri2018fundamental}. There we further develop a generic ``log-uniform'' search, characterized by a set of uninformative priors in which the probability distribution over mass is log-uniform. Fixing the $LR$ source properties (the equivalent pickup inductance $L_{\rm PU}$, resistance $R$, and receiver volume) and scaling out constants, eq. (\ref{eq:MN_value}) reduces to (eq. 170 of \cite{chaudhuri2018fundamental})
\begin{equation}\label{eq:MN_logvalue}
F_{\rm log}[S_{21}(\nu)]= \int_{\nu_{l}}^{\nu_{h}} d\nu\ \left( \frac{|S_{21}(\nu)|^{2}}{|S_{21}(\nu)|^{2}n(\nu) +1} \right)^{2}, 
\end{equation}
where $S_{21}(\nu)$ is the forward transmission through the matching network and $\nu_{l}$ and $\nu_{h}$ are the lower and upper limits of the search range. $n(\nu)=1/(\mathrm{exp}(h\nu/kT)-1)$ is the thermal occupancy of the signal source at frequency $\nu$. The ``+1'' term represents the noise at the SQL. 

We maximize $F_{\rm log}$. The signal source possesses complex impedance, and the quantum-limited amplifier possesses real noise impedance. If the matching network is linear, passive, and reciprocal, then the Bode-Fano criterion \cite{bode1945network,fano1950theoretical} applies,
yielding an upper bound on $F_{\rm log}$ (see Sec. V A 2 of \cite{chaudhuri2018fundamental}),
\begin{equation}\label{eq:BF_bound}
\frac{F_{\rm log}[S_{21}(\nu)]}{R/L_{\rm PU}} \lessapprox \begin{cases}
0.4, \hspace{1 cm} n(\nu_{h}) \ll 1 \\
\frac{1}{4n(\nu_{h})}, \hspace{0.5 cm} n(\nu_{h}) \gg 1
\end{cases}.
\end{equation}
The bound applies to an amplifier that measures forward power, as well as one that measures voltage or current (the ``op-amp mode'' \cite{clerk2010introduction}), e.g. a SQUID; see App. F of \cite{chaudhuri2018fundamental}. An analogous bound applies to an equivalent-$RC$ (resistor and capacitor) source. 

Equality in eq. (\ref{eq:BF_bound}) is obtained with a narrowband, multipole LC Chebyshev filter, but such structures are difficult to implement. Therefore, we ask how close a single-pole resonator can come to the Bode-Fano limit.

\begin{figure}[thp] 
\centering
\includegraphics[width=8.6cm]{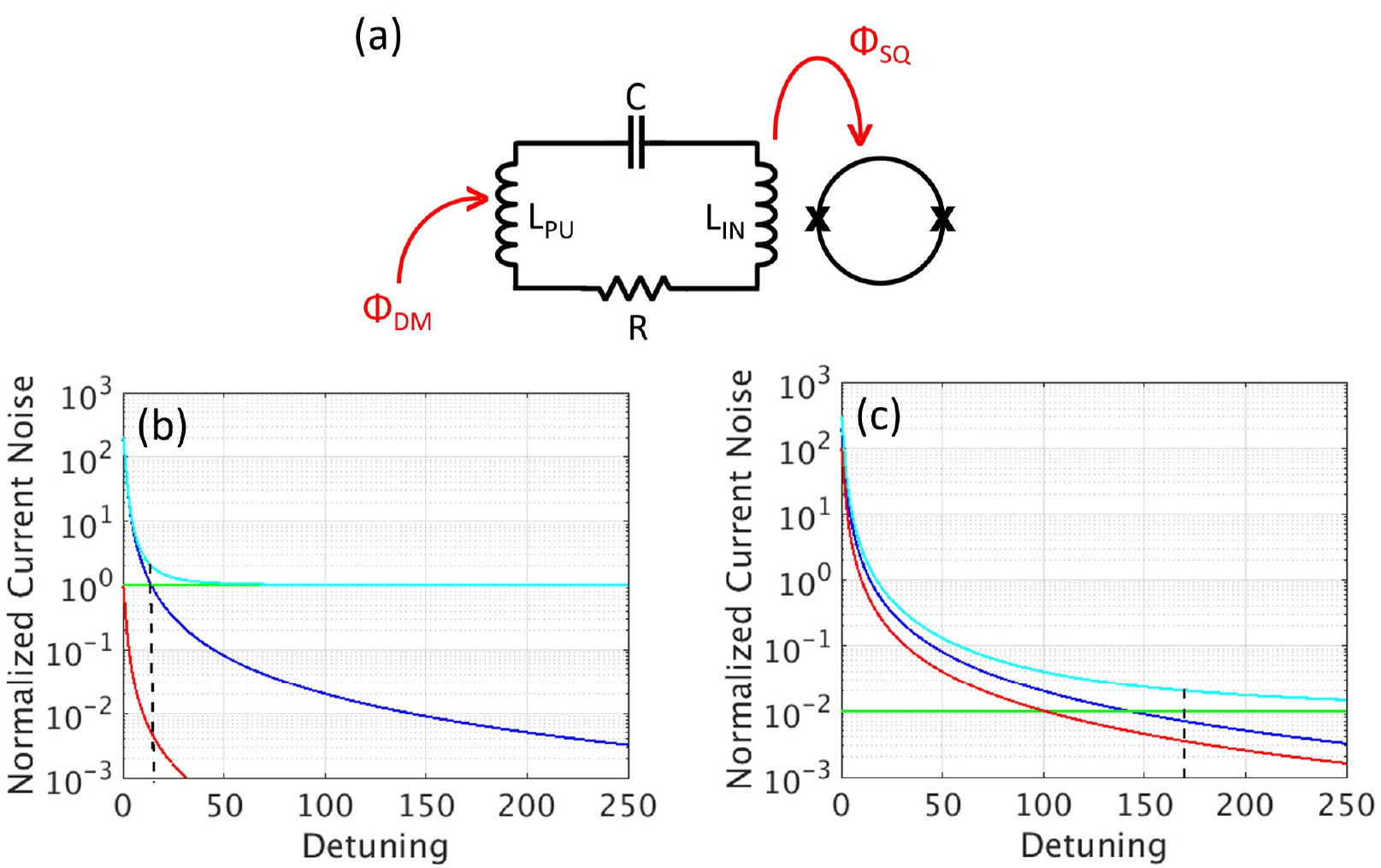}
\caption{(a) Equivalent-RLC search. The flux $\Phi_{DM}$ couples to the inductor, driving flux $\Phi_{SQ}$ in the amplifier, shown schematically as a dc SQUID. We plot detuning $2Q \frac{\nu-\nu_{r}}{\nu_{r}}$ vs. input-referred current noise spectral density $S_{II} /(h\nu_{r}/R)$ in the (b) noise-matched and (c) optimally-matched cases for $n(\nu_{r})=50$. Blue: thermal + zero-point noise, green: imprecision, red: backaction, cyan: total noise. Detunings below the dashed black line are in the sensitivity bandwidth. \label{fig:ResOpt}}
\end{figure}

Suppose, having already fixed the $LR$ source properties, we also fix the resonant frequency $\nu_{r}$. The receiver response is then determined by the strength of coupling to the amplifier, quantified by the ratio of noise impedance to signal source resistance. Optimizing eq. (\ref{eq:MN_logvalue}) with respect to this ratio, (Sec. V A 3 of \cite{chaudhuri2018fundamental})
\begin{equation}\label{eq:MN_logvalue_res}
\frac{F_{\rm log}^{\rm opt}(\nu_{r})}{R/L_{\rm PU}} \approx \begin{cases}
\frac{8}{27}, \hspace{1.4 cm} n(\nu_{r}) \ll 1 \\
\frac{1}{3\sqrt{3} n(\nu_{r})}, \hspace{0.5 cm} n(\nu_{r}) \gg 1
\end{cases}.
\end{equation}
Interestingly, we observe that, for optimal integrated sensitivity, the readout is noise-mismatched and dominated by backaction on resonance. To understand why, we consider readout with a quantum-limited flux-to-voltage amplifier (Fig. \ref{fig:ResOpt}(a)) and the contributions to the input-referred current noise for $n(\nu_{r})=50$. In the noise-matched case of Fig. \ref{fig:ResOpt}(b), the amplifier noise on resonance is minimized, with equal backaction and imprecision. The result is a sensitivity bandwidth (given approximately by the bandwidth for which imprecision is less than one-half of the total noise\cite{chaudhuri2018fundamental}) that is $2\sqrt{n(\nu_{r})} \approx 14$ resonator bandwidths. Now suppose that we couple more strongly to the amplifier, so that we achieve the optimal matching maximizing eq. (\ref{eq:MN_logvalue}). As illustrated in Fig. \ref{fig:ResOpt}(c), the backaction noise increases to one-half of the thermal noise on resonance, but the imprecision noise drops proportionally. In return for a 50\% SNR penalty near resonance, we achieve a much larger sensitivity bandwidth of $2n(\nu_{r})\sqrt{3} \approx 173$ resonator bandwidths, yielding a larger integrated sensitivity in eq. (\ref{eq:MN_logvalue}).

Taking the limit $\nu_{r} \rightarrow \nu_{h}$ in (\ref{eq:MN_logvalue_res}), we then find (eq. 203 of \cite{chaudhuri2018fundamental}) that the optimized single-pole resonator is approximately 75\% of the Bode-Fano limit (\ref{eq:BF_bound}). \emph{Single-pole resonators are near-optimal, single-moded detectors for a search over a wide band.} One consequence of this result is that, in contrast to the claims in \cite{kahn2016broadband}, a tunable equivalent-RLC circuit is superior to a broadband equivalent-LR circuit; see App. G of \cite{chaudhuri2018fundamental} for details.

We now return to a consideration of priors and discuss scanning. An optimal search requires scanning for complete coverage of a search band. Given a fixed total experiment time, what value function do we use to distribute time across scan steps? A scan consists of a set of resonant frequencies $\{\nu_{r}^{i} \}$; at each frequency, we integrate for time $\tau_{i}$. For axion searches, the time allocation is optimized by maximizing the weighted area of the excluded mass-coupling parameter space:
\begin{equation} \label{eq:A_axion}
A_{a}[ \{ \nu_{r}^{i} \} , \{ \tau_{i} \} ] = \int_{\nu_{l}}^{\nu_{h}} d\nu_{\rm DM}^{0} \int_{g_{a \gamma \gamma}^{\rm min}}^{g_{a\gamma \gamma}^{\rm max}} dg_{a\gamma \gamma}\ W_{a}.
\end{equation}
$W_{a}=W_{a}(\nu_{\rm DM}^{0}, g_{a\gamma \gamma})$ is the weighting function, based on mass and coupling priors introduced for eq. (\ref{eq:MN_value}). $g_{a\gamma\gamma}^{\rm min}=g_{a \gamma \gamma}^{\rm min}(\nu_{\rm DM}^{0}, \{ \nu_{r}^{i} \} , \{ \tau_{i} \})$ is the minimal coupling at $\nu_{\rm DM}^{0}$ to which the search is sensitive, based on the SNR accrued from integration at each scan step. $g_{a\gamma\gamma}^{\rm max}=g_{a\gamma \gamma}^{\rm max}(\nu_{\rm DM}^{0})$ is a maximal coupling, based on previous constraints. An analogous value function, using $\varepsilon$, can be defined for the hidden photon. For a log-uniform search, in which $W_{a}$ is proportional to the product of log-uniform distributions over mass and coupling, maximizing eq. (\ref{eq:A_axion}) shows that equal time should be spent in each octave of frequency space; see Sec. V B of \cite{chaudhuri2018fundamental}.



Our scan optimization also reveals that the SNR improves as the quality factor is increased above $Q_{\rm DM}=10^{6}$, the characteristic quality factor of the dark matter signal. To explain this result, we return to eq. (\ref{eq:Dicke}) and assume $\nu_{\rm DM}^{0}$ lies near resonance. In a single scan step with $Q>Q_{\rm DM}$, the signal power is roughly independent of $Q$ because an increase in $Q$ is compensated by a reduction in signal fraction within the resonator bandwidth. The noise power decreases as $1/Q$ because $\Delta \nu_{s}$ equals the resonator bandwidth. The integration time also decreases as $1/Q$ because more scan steps are required with a finer bandwidth. Combining all factors, the SNR from the single step is roughly independent of $Q$. Stepping at one part in $\sim Q$, we are sensitive to the same mass over $\sim Q/Q_{\rm DM}$ scan steps. The SNRs from different steps add in quadrature so the total SNR improves as $\sim \sqrt{Q}$.

\begin{figure}[htp] 
\includegraphics[width=8.6cm]{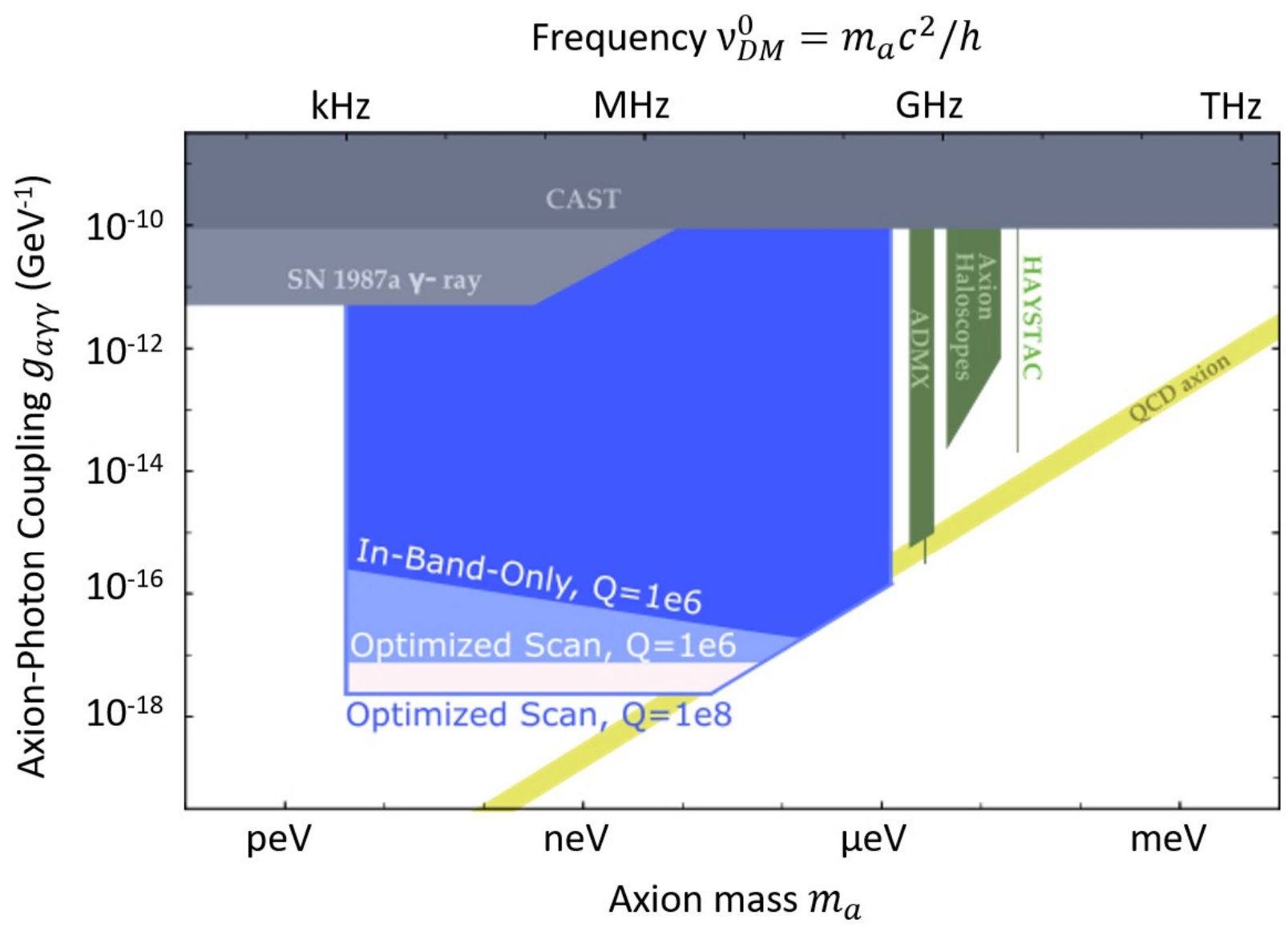}
\caption{Axion search with quantum-limited readout, 1 $m^{3}$ volume at 10 mK, 4 T magnetic field, and up to 100 days per e-folding integration time (stopped at DFSZ) over 1 kHz-300 MHz. Sensitivity (eq. 253 of \cite{chaudhuri2018fundamental}) for: a $Q=10^{6}$ scan using only information in the resonator bandwidth (dark blue), an optimized $Q=10^{6}$ scan (light blue), and an optimized $Q=10^{8}$ scan (pink). We have assumed a geometry factor of $c_{\rm PU}\approx 0.15$, which is reasonable for physically lumped-element receivers \cite{Chaudhuri:2014dla,Silva-Feaver:2016qhh}; see Section VI B of \cite{chaudhuri2018fundamental}. The QCD axion band (yellow) is bounded at the top and bottom by the KSVZ\cite{shifman1980can,kim1979weak} and DFSZ\cite{dine1981simple,zhitnitskij1980possible} axion models, respectively. \label{fig:LimitPlot}}. 
\end{figure}

In Fig. \ref{fig:LimitPlot}, we plot scan sensitivity to axion-photon coupling. A resonator can probe all frequencies within its sensitivity bandwidth without SNR degradation. Thus, relative to a noise-matched scan using only information in the resonator bandwidth (dark blue with $Q=10^{6}$), a scan with optimized integrated sensitivity (light blue with $Q=10^{6}$) increases the integration time at frequency $\nu_{\rm DM}^{0}$ by $\sim n(\nu_{\rm DM}^{0})$. Then, from eq. (\ref{eq:Dicke}), since signal power increases as coupling squared\cite{Sik85}, the minimum coupling to which the search is sensitive improves by $\sim n(\nu_{\rm DM}^{0})^{1/4}$. This corresponds to more than an order of magnitude increase in reach at $\sim$1 kHz, or equivalently, a five order of magnitude increase in scan rate. Additionally, sensitivity outside of the resonator bandwidth increases the step size needed for complete coverage of a search band, enabling significantly coarser tuning. By increasing $Q$ to $10^{8}$ (pink), we further enhance the reach by a factor of $\sqrt{10}$ (100$\times$ increase in scan rate). 

In summary, we have established a fundamental limit on electromagnetic searches for axion and hidden-photon dark matter (eq. \ref{eq:BF_bound}) including quantum noise from the amplifier. We have demonstrated that the single-pole resonator is a near-ideal single-moded technique for searches over a wide band, justifying the approach of multiple experiments \cite{du2018search,brubaker2017first,Chaudhuri:2014dla,Sik14}. We have shown that optimal scan strategies can increase scan rate by several orders of magnitude, particularly at low mass.



In light of these results, it is useful to ask how we may evade these quantum limits. One may evade the Bode-Fano constraint, for instance, by using a nonlinear, active, or multi-moded receiver. These receivers are often constrained by more generalized matching criteria \cite{youla1964new,nie2015broadband} and can be challenging to implement due to stability and parasitics. The other option is to improve on these limits by evading the SQL. Due to developments in quantum metrology, a number of techniques evading the SQL (e.g. squeezing, entanglement, backaction evasion, photon counting) can now be deployed.\cite{lehnert2016quantumaxion,backes2020quantum,dixit2020searching} Our results thus motivate the development of quantum measurement techniques in light-field dark matter searches.
\newline 


\begin{acknowledgments}
This research is funded in part by the Gordon and Betty Moore Foundation. Additional support was provided by the Heising-Simons Foundation, and some material is based upon work supported by the U.S. Department of Energy, Office of Science, National Quantum Information Science Research Centers. We thank Asimina Arvanitaki, Hsiao-Mei Cho, Carl Dawson, Michel Devoret, Stephen Kuenstner, Dale Li, Harvey Moseley, Lyman Page, Arran Phipps, Jamie Titus, Tony Tyson, Karl van Bibber, Betty Young, and Jonas Zmuidzinas for useful discussions.
\end{acknowledgments}

\bibliography{Bib_Opt}

\end{document}